# A Survey of Attacks, Security Mechanisms and Challenges in Wireless Sensor Networks

Dr. G. Padmavathi,

Prof and Head,
Dept. of Computer Science,
Avinashilingam University for Women,
Coimbatore, India,
ganapathi.padmavathi@gmail.com

Mrs. D. Shanmugapriya,

Lecturer,
Dept. of Information Technology,
Avinashilingam University for Women,
Coimbatore, India,
ds_priyaa@rediffmail.com

*Abstract*—**Wireless Sensor networks (WSN) is an emerging technology and have great potential to be employed in critical situations like battlefields and commercial applications such as building, traffic surveillance, habitat monitoring and smart homes and many more scenarios. One of the major challenges wireless sensor networks face today is security. While the deployment of sensor nodes in an unattended environment makes the networks vulnerable to a variety of potential attacks, the inherent power and memory limitations of sensor nodes makes conventional security solutions unfeasible. The sensing technology combined with processing power and wireless communication makes it profitable for being exploited in great quantity in future. The wireless communication technology also acquires various types of security threats. This paper discusses a wide variety of attacks in WSN and their classification mechanisms and different securities available to handle them including the challenges faced.**

*Keywords-Wireless Sensor Network; Security Goal; Security Attacks; Defensive mechanisms; Challenges*

## I. INTRODUCTION

Basically, sensor networks are application dependent. Sensor networks are primarily designed for real-time collection and analysis of low level data in hostile environments. For this reason they are well suited to a substantial amount of monitoring and surveillance applications. Popular wireless sensor network applications include wildlife monitoring, bushfire response, military command, intelligent communications, industrial quality control, observation of critical infrastructures, smart buildings, distributed robotics, traffic monitoring, examining human heart rates etc. Majority of the sensor network are deployed in hostile environments with active intelligent opposition. Hence security is a crucial issue. One obvious example is battlefield applications where there is a pressing need for secrecy of location and resistance to subversion and destruction of the network. Less obvious but just as important security dependent applications include:

- *Disasters:* In many disaster scenarios, especially those induced by terrorist activities, it may be necessary to protect the location of casualties from unauthorized disclosure

- *Public Safety:* In applications where chemical, biological or other environmental threats are monitored, it is vital that the availability of the network is never threatened. Attacks causing false alarms may lead to panic responses or even worse total disregard for the signals.

- *Home Healthcare:* In such applications, privacy protection is essential. Only authorized users should be able to query and monitor the network.

The major contribution of this paper includes classification of security attacks, security mechanisms and challenges in Wireless Sensor Networks. Section 2 gives the detailed information about the security goals in Wireless Sensor Networks. Security attacks and their classification are discussed in section 3. Section 4 discusses about the various security mechanisms. Major challenges faced are given in Section 5 followed by the conclusion section.

## II. SECURITY GOALS FOR SENSOR NETWORKS

As the sensor networks can also operate in an adhoc manner the security goals cover both those of the traditional networks and goals suited to the unique constraints of adhoc sensor networks. The security goals are classified as primary and secondary [5]. The primary goals are known as standard security goals such as Confidentiality, Integrity, Authentication and Availability (CIAA). The secondary goals are Data Freshness, Self-Organization, Time Synchronization and Secure Localization.

The primary goals are:

### A. Data Confidentiality

Confidentiality is the ability to conceal messages from a passive attacker so that any message communicated via the sensor network remains confidential. This is the most important issue in network security. A sensor node should not reveal its data to the neighbors.

## B. Data Authentication

Authentication ensures the reliability of the message by identifying its origin. Attacks in sensor networks do not just involve the alteration of packets; adversaries can also inject additional false packets [14]. Data authentication verifies the identity of the senders and receivers. Data authentication is achieved through symmetric or asymmetric mechanisms where sending and receiving nodes share secret keys. Due to the wireless nature of the media and the unattended nature of sensor networks, it is extremely challenging to ensure authentication.

## C. Data Integrity

Data integrity in sensor networks is needed to ensure the reliability of the data and refers to the ability to confirm that a message has not been tampered with, altered or changed. Even if the network has confidentiality measures, there is still a possibility that the data integrity has been compromised by alterations. The integrity of the network will be in trouble when:

- A malicious node present in the network injects false data.

- Unstable conditions due to wireless channel cause damage or loss of data.[4]

## D. Data Availability

Availability determines whether a node has the ability to use the resources and whether the network is available for the messages to communicate. However, failure of the base station or cluster leader's availability will eventually threaten the entire sensor network. Thus availability is of primary importance for maintaining an operational network.

The Secondary goals are:

## E. Data Freshness

Even if confidentiality and data integrity are assured, there is a need to ensure the freshness of each message. Informally, data freshness [4] suggests that the data is recent, and it ensures that no old messages have been replayed. To solve this problem a nonce, or another time-related counter, can be added into the packet to ensure data freshness.

## F. Self-Organization

A wireless sensor network is a typically an ad hoc network, which requires every sensor node be independent and flexible enough to be self-organizing and self-healing according to different situations. There is no fixed infrastructure available for the purpose of network management in a sensor network. This inherent feature brings a great challenge to wireless sensor network security. If self-organization is lacking in a sensor network, the damage resulting from an attack or even the risky environment may be devastating.

## G. Time Synchronization

Most sensor network applications rely on some form of time synchronization. Furthermore, sensors may wish to compute the end-to-end delay of a packet as it travels between two pairwise sensors. A more collaborative sensor network may require group synchronization [4] for tracking applications.

## H. Secure Localization

Often, the utility of a sensor network will rely on its ability to accurately and automatically locate each sensor in the network. A sensor network designed to locate faults will need accurate location information in order to pinpoint the location of a fault. Unfortunately, an attacker can easily manipulate nonsecured location information by reporting false signal strengths, replaying signals.

This Section has discussed about the security goals that are widely available for wireless sensor networks and the next section explains about the attacks that commonly occur on wireless sensor networks.

## III. ATTACKS ON SENSOR NETWORKS

Wireless Sensor networks are vulnerable to security attacks due to the broadcast nature of the transmission medium. Furthermore, wireless sensor networks have an additional vulnerability because nodes are often placed in a hostile or dangerous environment where they are not physically protected. Basically attacks are classified as active attacks and passive attacks. Figure1 shows the classification of attacks under general categories and Figure 2 shows the attacks classification on WSN.

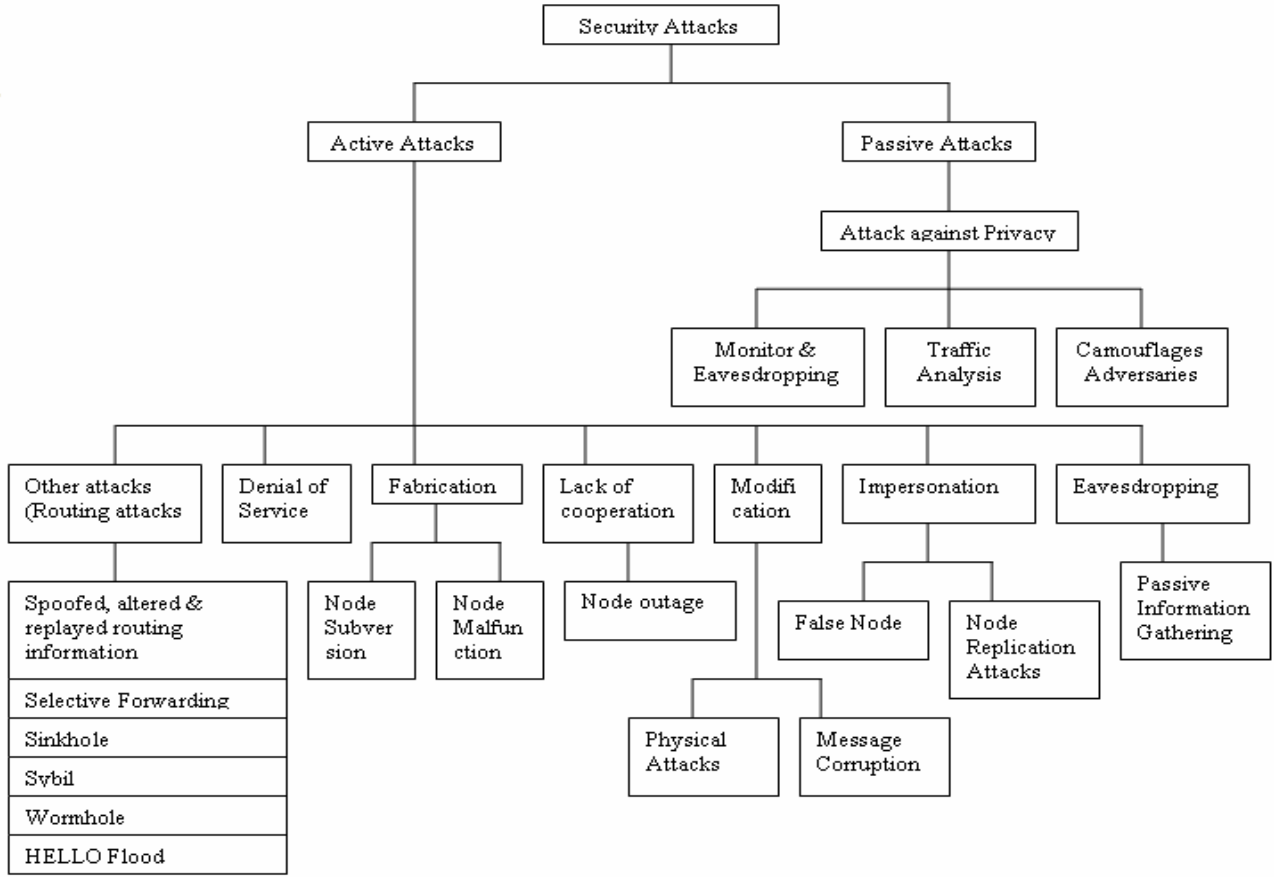

Figure 1.   General Classification of Security Attacks

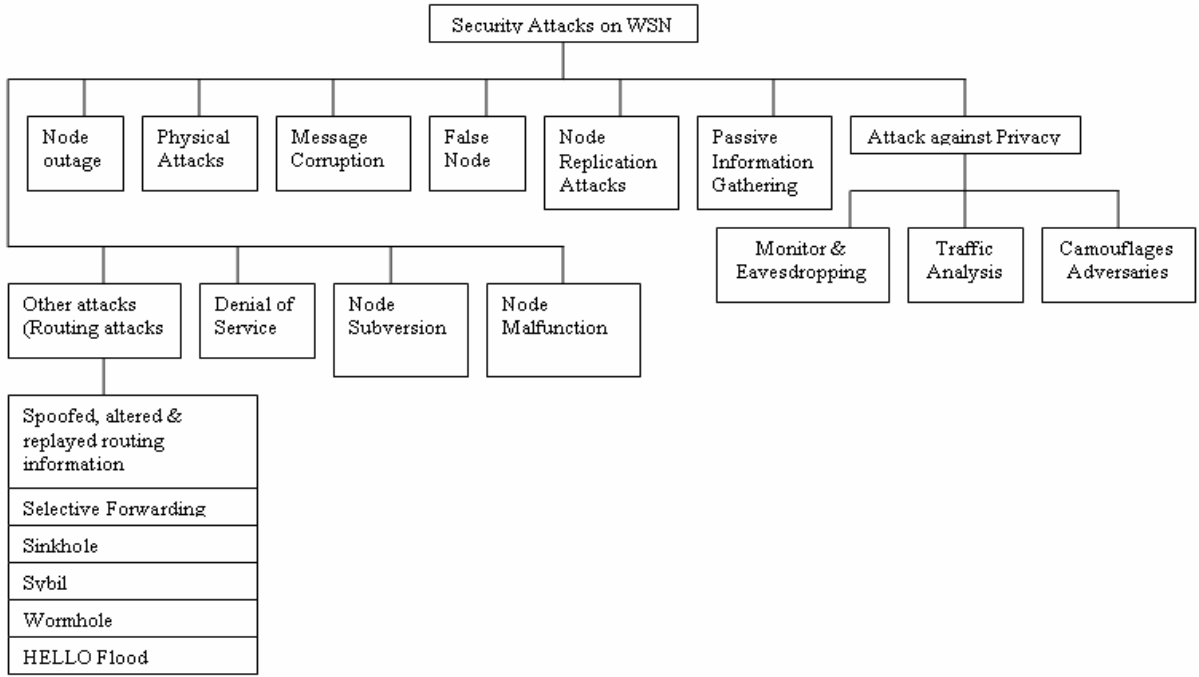

Figure 2.   Classification of Security Attacks on WSN

### A. Passive Attacks

The monitoring and listening of the communication channel by unauthorized attackers are known as passive attack. The Attacks against privacy is passive in nature.

#### 1) Attacks against Privacy

The main privacy problem is not that sensor networks enable the collection of information. In fact, much information from sensor networks could probably be collected through direct site surveillance. Rather, sensor networks intensify the privacy problem because they make large volumes of information easily available through remote access. Hence, adversaries need not be physically present to maintain surveillance. They can gather information at low-risk in anonymous manner. Some of the more common attacks[8] against sensor privacy are:

- *Monitor and Eavesdropping:* This is the most common attack to privacy. By snooping to the data, the adversary could easily discover the communication contents. When the traffic conveys the control information about the sensor network configuration, which contains potentially more detailed information than accessible through the location server, the eavesdropping can act effectively against the privacy protection.

- *Traffic Analysis:* Even when the messages transferred are encrypted, it still leaves a high possibility analysis of the communication patterns. Sensor activities can potentially reveal enough information to enable an adversary to cause malicious harm to the sensor network.

- *Camouflage Adversaries:* One can insert their node or compromise the nodes to hide in the sensor network. After that these nodes can copy as a normal node to attract the packets, then misroute the packets, conducting the privacy analysis.

### B. Active Attacks

The unauthorized attackers monitors, listens to and modifies the data stream in the communication channel are known as active attack. The following attacks are active in nature.

1. Routing Attacks in Sensor Networks
2. Denial of Service Attacks
3. Node Subversion
4. Node Malfunction
5. Node Outage
6. Physical Attacks
7. Message Corruption
8. False Node
9. Node Replication Attacks
10. Passive Information Gathering

#### 1) Routing Attacks in Sensor Networks

The attacks which act on the network layer are called routing attacks. The following are the attacks that happen while routing the messages.

##### a) Spoofed, altered and replayed routing information

- An unprotected ad hoc routing is vulnerable to these types of attacks, as every node acts as a router, and can therefore directly affect routing information.

- Create routing loops

- Extend or shorten service routes

- Generate false error messages

- Increase end-to-end latency [3]

##### b) Selective Forwarding

A malicious node can selectively drop only certain packets. Especially effective if combined with an attack that gathers much traffic via the node. In sensor networks it is assumed that nodes faithfully forward received messages. But some compromised node might refuse to forward packets, however neighbors might start using another route.[3]

##### c) Sinkhole Attack

Attracting traffic to a specific node in called sinkhole attack. In this attack, the adversary's goal is to attract nearly all the traffic from a particular area through a compromised node. Sinkhole attacks typically work by making a compromised node look especially attractive to surrounding nodes. [3]

##### d) Sybil Attacks

A single node duplicates itself and presented in the multiple locations. The Sybil attack targets fault tolerant schemes such as distributed storage, multipath routing and topology maintenance. In a Sybil attack, a single node presents multiple identities to other nodes in the network. Authentication and encryption techniques can prevent an outsider to launch a Sybil attack on the sensor network.[3]

##### e) Wormholes Attacks

In the wormhole attack, an attacker records packets (or bits) at one location in the network, tunnels them to another location, and retransmits them into the network.[3]

##### f) HELLO flood attacks

An attacker sends or replays a routing protocol's HELLO packets from one node to another with more energy. This attack uses HELLO packets as a weapon to convince the sensors in WSN. In this type of attack an attacker with a high radio transmission range and processing power sends HELLO packets to a number of sensor nodes that are isolated in a large area within a WSN. The sensors are thus influenced that the adversary is their neighbor. As a result, while sending the information to the base station, the victim nodes try to go through the attacker as they know that it is their neighbor and are ultimately spoofed by the attacker.[3]

### 2) Denial of Service

Denial of Service (DoS) is produced by the unintentional failure of nodes or malicious action. DoS attack is meant not only for the adversary's attempt to subvert, disrupt, or destroy a network, but also for any event that diminishes a network's capability to provide a service. In wireless sensor networks, several types of DoS attacks in different layers might be performed. At physical layer the DoS attacks could be jamming and tampering, at link layer, collision, exhaustion and unfairness, at network layer, neglect and greed, homing, misdirection, black holes and at transport layer this attack could be performed by malicious flooding and de-synchronization. The mechanisms to prevent DoS attacks include payment for network resources, pushback, strong authentication and identification of traffic.[2]

### 3) Node Subversion

Capture of a node may reveal its information including disclosure of cryptographic keys and thus compromise the whole sensor network. A particular sensor might be captured, and information (key) stored on it might be obtained by an adversary. [6]

### 4) Node Malfunction

A malfunctioning node will generate inaccurate data that could expose the integrity of sensor network especially if it is a data-aggregating node such as a cluster leader [6].

### 5) Node Outage

Node outage is the situation that occurs when a node stops its function. In the case where a cluster leader stops functioning, the sensor network protocols should be robust enough to mitigate the effects of node outages by providing an alternate route [6].

### 6) Physical Attacks

Sensor networks typically operate in hostile outdoor environments. In such environments, the small form factor of the sensors, coupled with the unattended and distributed nature of their deployment make them highly susceptible to physical attacks, i.e., threats due to physical node destructions. Unlike many other attacks mentioned above, physical attacks destroy sensors permanently, so the losses are irreversible. For instance, attackers can extract cryptographic secrets, tamper with the associated circuitry, modify programming in the sensors, or replace them with malicious sensors under the control of the attacker.

### 7) Message Corruption

Any modification of the content of a message by an attacker compromises its integrity.[9]

### 8) False Node

A false node involves the addition of a node by an adversary and causes the injection of malicious data. An intruder might add a node to the system that feeds false data or prevents the passage of true data. Insertion of malicious node is one of the most dangerous attacks that can occur. Malicious code injected in the network could spread to all nodes, potentially destroying the whole network, or even worse, taking over the network on behalf of an adversary.[9]

### 9) Node Replication Attacks

Conceptually, a node replication attack is quite simple; an attacker seeks to add a node to an existing sensor network by copying the nodeID of an existing sensor node. A node replicated in this approach can severely disrupt a sensor network's performance. Packets can be corrupted or even misrouted. This can result in a disconnected network, false sensor readings, etc. If an attacker can gain physical access to the entire network he can copy cryptographic keys to the replicated sensor nodes. By inserting the replicated nodes at specific network points, the attacker could easily manipulate a specific segment of the network, perhaps by disconnecting it altogether.[1]

### 10) Passive Information Gathering

An adversary with powerful resources can collect information from the sensor networks if it is not encrypted. An intruder with an appropriately powerful receiver and well-designed antenna can easily pick off the data stream. Interception of the messages containing the physical locations of sensor nodes allows an attacker to locate the nodes and destroy them. Besides the locations of sensor nodes, an adversary can observe the application specific content of messages including message IDs, timestamps and other fields. To minimize the threats of passive information gathering, strong encryption techniques needs to be used.[8]

This section explained about the attacks and their classification that widely happens on wireless sensor networks. The next section discusses about the security mechanisms that are used to handle the attacks.

## IV.  SECURITY MECHANISM

The security mechanisms are actually used to detect, prevent and recover from the security attacks. A wide variety of security schemes can be invented to counter malicious attacks and these can be categorized as high-level and low-level. Figure 3 shows the order of security mechanisms.

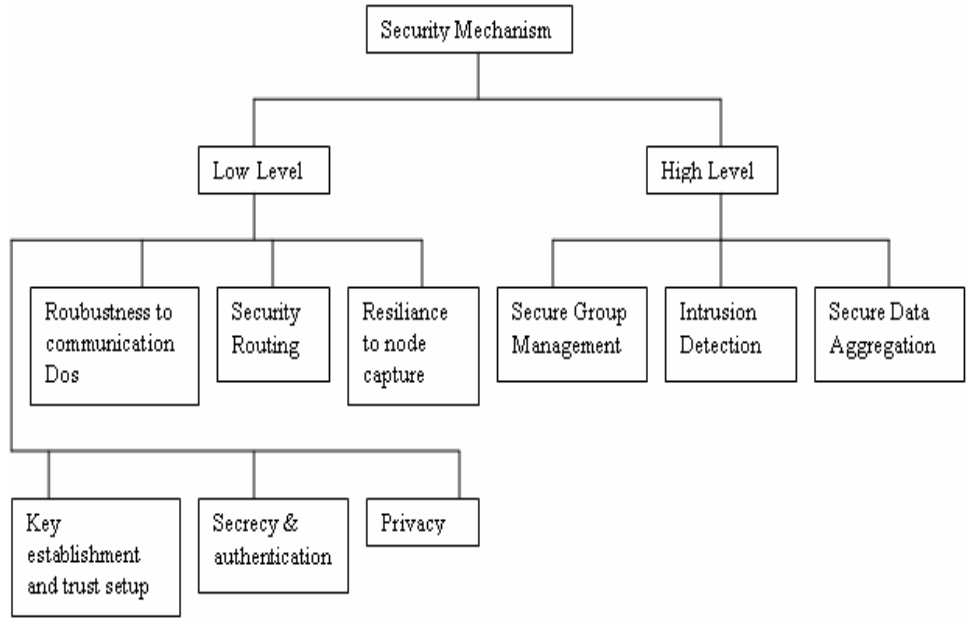

**Figure3: Security mechanisms**

### A. Low-Level Mechanism

Low-level security primitives for securing sensor networks includes,

1. Key establishment and trust setup
2. Secrecy and authentication
3. Privacy
4. Robustness to communication denial of service
5. Secure routing
6. Resilience to node capture

#### 1) Key establishment and trust setup

The primary requirement of setting up the sensor network is the establishment of cryptographic keys. Generally the sensor devices have limited computational power and the public key cryptographic primitives are too expensive to follow. Key-establishment techniques need to scale to networks with hundreds or thousands of nodes. In addition, the communication patterns of sensor networks differ from traditional networks; sensor nodes may need to set up keys with their neighbors and with data aggregation nodes. The disadvantage of this approach is that attackers who compromised sufficiently and many nodes could also reconstruct the complete key pool and break the scheme.[1]

#### 2) Secrecy and authentication.

Most of the sensor network applications require protection against eavesdropping, injection, and modification of packets. Cryptography is the standard defense. Remarkable system trade-offs arise when incorporating cryptography into sensor networks. For point-to-point communication[12], end-to-end cryptography achieves a high level of security but requires that keys be set up among all end points and be incompatible with passive participation and local broadcast. Link-layer cryptography with a network wide shared key simplifies key setup and supports passive participation and local broadcast, but intermediate nodes might eavesdrop or alter messages. The earliest sensor networks are likely to use link layer cryptography, because this approach provides the greatest ease of deployment among currently available network cryptographic approaches.[6]

#### 3) Privacy

Like other traditional networks, the sensor networks have also force privacy concerns. Initially the sensor networks are deployed for legitimate purpose might subsequently be used in unanticipated ways. Providing awareness of the presence of sensor nodes and data acquisition is particularly important. [1]

#### 4) Robustness to communication denial of service

An adversary attempts to disrupt the network's operation by broadcasting a high-energy signal. If the transmission is powerful enough, the entire system's communication could be jammed. More sophisticated attacks are also possible; the adversary might inhibit communication by violating the 802.11 medium access control (MAC) protocol by, say, transmitting while a neighbor is also transmitting or by continuously requesting channel access with a request-to-send signal.[1]

#### 5) Secure routing

Routing and data forwarding is a crucial service for enabling communication in sensor networks. Unfortunately, current routing protocols suffer from many security vulnerabilities. For example, an attacker might launch denial-of-service attacks on the routing protocol, preventing communication. The simplest attacks involve injecting malicious routing information into the network, resulting in routing inconsistencies. Simple authentication might guard against injection attacks, but some routing protocols are susceptible to replay by the attacker of legitimate routing messages. [6]

### 6) Resilience to node capture

One of the most challenging issues in sensor networks is resiliency against node capture attacks. In most applications, sensor nodes are likely to be placed in locations easily accessible to attackers. Such exposure raises the possibility that an attacker might capture sensor nodes, extract cryptographic secrets, modify their programming, or replace them with malicious nodes under the control of the attacker. Tamper-resistant packaging may be one defense, but it's expensive, since current technology does not provide a high level of security. Algorithmic solutions to the problem of node capture are preferable.[1]

### B. High-Level Mechanism

High-level security mechanisms for securing sensor networks, includes secure group management, intrusion detection, and secure data aggregation.

### 1) Secure group management

Each and every node in a wireless sensor network is limited in its computing and communication capabilities. However, interesting in-network data aggregation and analysis can be performed by groups of nodes. For example, a group of nodes might be responsible for jointly tracking a vehicle through the network. The actual nodes comprising the group may change continuously and quickly. Many other key services in wireless sensor networks are also performed by groups. Consequently, secure protocols for group management are required, securely admitting new group members and supporting secure group communication. The outcome of the group key computation is normally transmitted to a base station. The output must be authenticated to ensure it comes from a valid group. [1]

### 2) Intrusion detection

Wireless sensor networks are susceptible to many forms of intrusion. Wireless sensor networks require a solution that is fully distributed and inexpensive in terms of communication, energy, and memory requirements. The use of secure groups may be a promising approach for decentralized intrusion detection.[1]

### 3) Secure data aggregation

One advantage of a wireless sensor network is the fine-grain sensing that large and dense sets of nodes can provide. The sensed values must be aggregated to avoid overwhelming amounts of traffic back to the base station. For example, the system may average the temperature of a geographic region, combine sensor values to compute the location and velocity of a moving object, or aggregate data to avoid false alarms in real-world event detection. Depending on the architecture of the wireless sensor network, aggregation may take place in many places in the network. All aggregation locations must be secured.[6]

## V. CHALLENGES OF SENSOR NETWORKS

The nature of large, ad-hoc, wireless sensor networks presents significant challenges in designing security schemes. A wireless sensor network is a special network which has many constraint compared to a traditional computer network.

### A. Wireless Medium

The wireless medium is inherently less secure because its broadcast nature makes eavesdropping simple. Any transmission can easily be intercepted, altered, or replayed by an adversary. The wireless medium allows an attacker to easily intercept valid packets and easily inject malicious ones. Although this problem is not unique to sensor networks, traditional solutions must be adapted to efficiently execute on sensor networks. [7]

### B. Ad-Hoc Deployment

The ad-hoc nature of sensor networks means no structure can be statically defined. The network topology is always subject to changes due to node failure, addition, or mobility. Nodes may be deployed by airdrop, so nothing is known of the topology prior to deployment. Since nodes may fail or be replaced the network must support self-configuration. Security schemes must be able to operate within this dynamic environment.

### C. Hostile Environment

The next challenging factor is the hostile environment in which sensor nodes function. Motes face the possibility of destruction or capture by attackers. Since nodes may be in a hostile environment, attackers can easily gain physical access to the devices. Attackers may capture a node, physically disassemble it, and extract from it valuable information (e.g. cryptographic keys). The highly hostile environment represents a serious challenge for security researchers.

### D. Resource Scarcity

The extreme resource limitations of sensor devices pose considerable challenges to resource-hungry security mechanisms. The hardware constraints necessitate extremely efficient security algorithms in terms of bandwidth, computational complexity, and memory. This is no trivial task. Energy is the most precious resource for sensor networks. Communication is especially expensive in terms of power. Clearly, security mechanisms must give special effort to be communication efficient in order to be energy efficient. [5]

### E. Immense Scale

The proposed scale of sensor networks poses a significant challenge for security mechanisms. Simply networking tens to hundreds of thousands of nodes has proven to be a substantial task. Providing security over such a network is equally challenging. Security mechanisms must be scalable to very large networks while maintaining high computation and communication efficiency.

*F.  Unreliable Communication*

Certainly, unreliable communication is another threat to sensor security. The security of the network relies heavily on a defined protocol, which in turn depends on communication.[5]

- *Unreliable Transfer:* Normally the packet-based routing of the sensor network is connectionless and thus inherently unreliable.

- *Conflicts:* Even if the channel is reliable, the communication may still be unreliable. This is due to the broadcast nature of the wireless sensor network.

- *Latency:* The multi-hop routing, network congestion and node processing can lead to greater latency in the network, thus making it difficult to achieve synchronization among sensor nodes.

*G.  Unattended Operation*

Depending on the function of the particular sensor network, the sensor nodes may be left unattended for long periods of time. There are three main cautions to unattended sensor nodes [5]:

- *Exposure to Physical Attacks:* The sensor may be deployed in an environment open to adversaries, bad weather, and so on. The probability that a sensor suffers a physical attack in such an environment is therefore much higher than the typical PCs, which is located in a secure place and mainly faces attacks from a network.

- *Managed Remotely:* Remote management of a sensor network makes it virtually impossible to detect physical tampering and physical maintenance issues.

- *No Central Management Point:* A sensor network should be a distributed network without a central management point. This will increase the vitality of the sensor network. However, if designed incorrectly, it will make the network organization difficult, inefficient, and fragile.

Perhaps most importantly, the longer that a sensor is left unattended the more likely that an adversary has compromised the node.

## VI.  CONCLUSION

The deployment of sensor nodes in an unattended environment makes the networks vulnerable. Wireless sensor networks are increasingly being used in military, environmental, health and commercial applications. Sensor networks are inherently different from traditional wired networks as well as wireless ad-hoc networks. Security is an important feature for the deployment of Wireless Sensor Networks. This paper summarizes the attacks and their classifications in wireless sensor networks and also an attempt has been made to explore the security mechanism widely used to handle those attacks. The challenges of Wireless Sensor Networks are also briefly discussed. This survey will hopefully motivate future researchers to come up with smarter and more robust security mechanisms and make their network safer.

AUTHORS PROFILE

**Dr. Padmavathi Ganapathi** is the Professor and Head of Department of Computer Science, Avinashilingam University for Women, Coimbatore. She has 21 years of teaching experience and one year Industrial experience. Her areas of interest include Network security and Cryptography and real time communication. She has more than 50 publications at national and International level. She is a life member of many professional organizations like CSI, ISTE, AACE, WSEAS, ISCA, and UWA.

**Mrs. Shanmugapriya. D,** received the B.Sc. and M.Sc. degrees in Computer Science from Avinashilingam University for Women, Coimbatore in 1999 and 2001 respectively. And, she received the M.Phil degree in Computer Science from Manonmaniam Sundaranar University, Thirunelveli in 2003 and pursuing her PhD at Avinashilingam University for Women. She is currently working as a Lecturer in Information Technology in the same University and has eight years of teaching experience. Her research interests are Biometrics, Network Security and System Security.